\documentclass{PoS}

\usepackage{slashed} 
\usepackage{caption} 
\usepackage{nccmath} 
\usepackage{times} 
\usepackage{amsmath} 
\usepackage{amssymb} 
\usepackage{verbatim} 
\usepackage{graphicx} 
\usepackage{color} 
\usepackage{booktabs} 
\usepackage{subfigure} 
\newcommand{\lp}{\left(} \newcommand{\rp}{\right)}

\title{Strong field effects on physics processes at the Interaction Point of future linear colliders}

\ShortTitle{Strong field effects at the IP}

\author{\speaker{A. Hartin}\\
        DESY, Hamburg\\
        E-mail: \email{anthony.hartin@desy.de}}

\author{G. Moortgat-Pick\\
        DESY/II. Institute of theoretical physics, University of Hamburg\\
        }
\author{S. Porto\\
        II. Institute of theoretical physics, University of Hamburg\\
       }

\abstract{Future lepton colliders will be precision machines whose physics program includes close study of the Higgs sector and searches for new physics via polarised beams. The luminosity requirements of such machines entail very intense lepton bunches at the interaction point with associated strong electromagnetic fields. These strong fields not only lead to obvious phenomena such as beamstrahlung, but also potentially affect every particle physics process via virtual exchange with the bunch fields. For precision studies, strong field effects have to be understood to the sub-percent level. Strong external field effects can be taken into account exactly via the Furry Picture or, in certain limits, via the Quasi-classical Operator method . Significant theoretical development is in progress and here we outline the current state of play.}

\FullConference{36th International Conference on High Energy Physics,\\
		July 4-11, 2012\\
		Melbourne, Australia}

\begin{document}

\section{Introduction}

Quantum tunnelling is a well known and utilised phenomena that involves the interaction of charged particles with a potential barrier. The quantum tunnelling of an electron exists only if the width of the potential barrier is less  than a Compton wavelength and it can be interpreted as an induced pair creation by an over-critical potential \cite{Nikish70, Greiner85}. Interpreted this way, quantum tunnelling is related to the Schwinger effect in which an external electromagnetic field creates a real electron positron pair from the vacuum. The (Schwinger) critical field strength required ($1.3\times10^{18}$ V/m) must be sufficient to induce an energy difference of two rest masses across a Compton wavelength. \\ 

The Schwinger critical field strength requires extreme physical conditions in order to produce. It may be realised near the surface of astronomical objects \cite{KuzMik04}, in the rest frame of relativistic particles by a new generation of ultra-intense lasers \cite{ELIgrand2009}, in heavy ion collisions \cite{Baur07} and in future linear colliders (the International Linear Collider (ILC) and the Compact Linear Collider (CLIC)). \\

Strong electromagnetic fields can be considered as classical objects rather than ensembles of quantised particles due to a high number density of associated quanta and the overlap of their wavefunctions. Additionally, equations of motion of quantised charged particles minimally coupled to external electromagnetic fields can be solved exactly \cite{Volkov35}. These exact solutions can be included in the Quantum Field Theory via the Furry Picture (FP) in the Feynman-Schwinger-Tomonaga formulation via the S-matrix theory \cite{Furry51} or via the path integral approach \cite{VaiFarHot92}. \\

An important issue for implementing the FP is the nature of the vacuum at or beyond the critical field. In such a regime the vacuum can become charged \cite{Greiner85}. Consequently charge parity is lost, the vacuum becomes unstable and tadpole diagrams are non-vanishing \cite{FraGitShv91}. The unstable vacuum can be discounted for incoming and outgoing states in S-matrix theory since they operate on the vacuum at $t=\pm\infty$ at which, in most circumstances, the external field adiabatically vanishes.  \\

In this paper we concentrate on the intense electromagnetic fields, associated with tightly squeezed charge bunches colliding at the interaction point of future linear colliders. We establish the types of fields expected there, the scale of field strengths expected and the requirements for calculation of the probability of physics processes in such fields within the FP. An alternative approach useful for ultra-relativistic particles, the quasi-classical operator method, will also be discussed.

\section{The Furry picture}

Within the usual perturbation theory, the interaction of an initial particle with an intense field considered to be composed of many identical photons, involves a redundancy. That is, there are many ways in which photons can be exchanged with the field in order to produce a particular final state. It turns out for the case of oscillatory fields at least that the normal coupling constant of the perturbation theory is multiplied by a factor related to the intensity of the field \cite{FriEbe64,Reiss05}.\\

Therefore for particle interactions with intense external fields it is desirable to rework perturbation theory in order to take into account the external field exactly. Such can be achieved by using the FP which separates the external field $A^\text{ext}$ from the quantised boson field in the interaction Lagrangian. For example the QED Lagrangian and the fermion equation of motion in the FP are

\begin{gather}\label{eqn1}
\mathcal{L_{\text{QED}}^{\text{FP}}}\equiv\bar\psi^{\text{FP}}(i\slashed{\partial}-e\slashed{A}^{\text{ext}}-m)\psi^{\text{FP}}-\frac{1}{4}(F_{\mu\nu})^2-e\bar\psi^{\text{FP}}\slashed{A}\,\psi^{\text{FP}} \notag \\
\lp i\slashed{\partial}-e\slashed{A}^{\text{ext}}-m \rp \psi^{\text{FP}}=0 
\end{gather}

Specific solutions of the Dirac equation minimally coupled to the external field are required. These have been obtained for plane wave fields, longitudinal fields and some collinear combinations of the two \cite{BagGit90,Lyulka75,Hartin11}. In the limit of vanishing external field the FP reduces to the usual interaction picture. Therefore for physics processes taking place in external fields, such as those at the interaction point of a collider, the FP is the more complete description. In order to employ the FP in collider processes we need the specific external field configuration and new solutions of the FP equations of motion.

\section{Strong field collider processes in the Furry picture}

Particle physics experiments in particle colliders involve the overlap of intense bunches of charged particles which interact to produce final states of interest. The treatment of collective effects from the strong electromagnetic fields at tree level has been limited to the treatment of beamstrahlung and its related process (via a crossing symmetry), coherent pair production.\\

The beamstrahlung has been studied in the FP \cite{Ritus72} and via an alternative strong field theory - the quasi-classical operator method (see section \ref{sect4}). The transition probabilities depend on the external field through a parameter $\Upsilon$ which is a ratio of the field strength experienced by an incident charged particle of momentum $p$, mass $m$ and charge $e$, to the Schwinger critical field strength, 

\begin{gather}\label{eqn3}
\Upsilon=\frac{\sqrt{(eF_{\mu\nu}p_\nu)^2}}{m^3} 
\end{gather}

The $\Upsilon$ parameter for a particular collider depends on the interaction point beam parameters - the bunch population N, the bunch dimensions $\sigma_x,\sigma_y,\sigma_z$ and the bunch particle energy $E$ and mass $m$ (expressed as $\gamma=E/m$) - as well as the Compton radius $r_e$ and the fine structure constant $\alpha$. For a real bunch collision the $\Upsilon$ parameter varies due to bunch disruption and the pinch effect. The average $\Upsilon_{\text{av}}$ is \cite{YokChe91},

\begin{gather}\label{eqn4}
\Upsilon_{\text{av}}=\mfrac{5}{6}\mfrac{N\;\gamma\; r_e^2}{\alpha (\sigma_x+\sigma_y)\sigma_z}
\end{gather}

\begin{table}[h!]\centering\begin{tabular}{|c|| c | c | c | c |}\hline
 Machine & LEP2 & SLC & ILC & CLIC \\ \hline\hline
   E (GeV) & 94.5 & 46.6 & 500 & 1500 \\ \hline
    $N(\times 10^{10})$ & $334 $ & $4$ & $ 2 $ & $ 0.37 $\\ \hline
   $\sigma_x,\sigma_y$ ($\mu$m) & 190, 3 & 2.1, 0.9 & 0.49, 0.002 & 0.045, 0.001\\ \hline
   $\sigma_z $ (mm) & 20 & 1.1  & 0.15 & 0.044\\ \hline
   $\Upsilon_{\text{av}}$ & 0.00015 & 0.001 & 0.24 & 4.9 \\ \hline
\end{tabular}\caption{\bf collider beam parameters and the strong field $\Upsilon_{\text{av}}$}\label{table1}\end{table}

Historically the $\Upsilon$ associated with the beam parameters of a particular collider, has increased over the last decades as colliders have become more energetic and charge bunches have become more compressed (table \ref{table1}). Of critical importance is how FP process transition probabilities scale with increasing $\Upsilon$. \\

The beamstrahlung and coherent pair production, being 1-vertex processes in the FP, do not occur for mass-shell states in the normal interaction picture. So in the FP the beamstrahlung and coherent pair production transition probabilities go to zero as $\Upsilon$ vanishes. In contrast, in the extreme strong field limit $\Upsilon\gg 1$ the beamstrahlung transition probability scales as $\Upsilon^{2/3}$ \cite{Ritus72}. \\

In general, however, the transition probability dependence on $\Upsilon$ is complicated and specific calculations are required to establish numerical values. There is no a priori argument then as to why higher order FP processes should be neglected in the formulation of physics process transition probabilities in the new generation of colliders. \\

The beamstrahlung and coherent pair production, as currently formulated and simulated, assume interaction with only one of strong fields present at the interaction point - that of the oncoming bunch to the incident particle. For a general collider physics process, involving propagators and integration over all possible final momenta, the strong fields of {\bf both} interacting charge bunches must be taken into account. \\ 

In order to employ the FP in say, future linear colliders, solutions of the equations of motion of all particles coupled to the overlapping fields are required. This work is in progress and once these solutions once found corrections to {\bf all} collider processes will be calculated. 

\section{The quasi-classical operator method}\label{sect4}

An alternative, albeit less general, theory to the FP, which incorporates strong external fields, is the quasi-classical operator method (QOM). The QOM makes use of the fact that particle physics experiments often involve ultra-relativistic particles so that the commutator of their dynamical motion is very small. Setting that commutator to zero allows the particle motion to be considered classical. This is the quasi-classical approximation. \\

The QOM is a quantum theory within the Heisenberg picture and utilises operator calculus until a late stage in the calculation of a transition probability. The QOM has been applied to calculate beamstrahlung and coherent pair production in colliders, as well as to crystals containing strong inter-lattice fields \cite{Baier09}. \\

It has been shown, for the case of the beamstrahlung at least, that the transition probabilities obtained using the FP and the QOM are asymptotically identical in the ultra-relativistic limit \cite{Hartin09}. Though in general the FP does not require an ultra-relativistic limit and applies to all particle momenta. It may be possible to apply the QOM to additional collider processes by using the quasi-classical approximation for ultra-relativistic initial states, selecting intermediate states which don't couple to the external fields and relying on the suppression of the $\Upsilon$ parameter for heavy final states. An investigation into such an application is being undertaken by our group.

\section{Conclusion}

Many new physics experiments which are being planned and/or are under construction, will involve physics processes in external fields which approach or exceed the Schwinger critical field. In such a strong field environment, the coupling constant of the normal perturbation expansion becomes dependent on the external field intensity and new theories which take into account the interaction with the external field {\bf exactly} are required. \\

The FP, which separates the external field from the interaction Lagrangian and requires solutions of the minimally coupled equations of motion is such a theory. The QOM is an alternative theory in the Heisenberg picture which relies on the limiting case of ultra-relativistic particles. \\

We seek to apply these theories to collider phenomenology for future linear colliders in which two strong non-collinear fields (those of the colliding charge bunches) are present at the interaction point. The FP requires new solutions of the equations of motion, external field propagators and radiative corrections, in the two non-collinear electromagnetic fields. The FP applies to all physics processes taking place at the interaction point. The QOM can be applied to a subset of processes in which the quasi-classical approximation is valid. \\

In previous lepton colliders, the bunch fields were sufficiently small so that strong field effects beyond the first order in the FP perturbation expansion (i.e. beyond the beamstrahlung) could be neglected. However in the future generation of colliders the field is potentially strong enough to noticeably affect the phenomenology. This provides the motivation for strong field theoretical study described in this paper.


\end{document}